\documentclass[%
 reprint,
superscriptaddress,https://www.overleaf.com/project/64994f20bd1198e1d7691150
 amsmath,amssymb,
 aps,pre,hyphens,floatfix
]{revtex4-2}

\usepackage{lineno}

\usepackage{hyperref}
\hypersetup{hypertex=true,
colorlinks=true,
linkcolor=blue,
anchorcolor=blue,
citecolor=blue}

\usepackage{amsfonts}
\usepackage{amsmath}
\usepackage[]{amsmath}
\usepackage{amssymb}
\usepackage{multirow} 
\usepackage{enumerate}
\usepackage[usenames,dvipsnames,svgnames,table]{xcolor}
\usepackage{IEEEtrantools}
\usepackage{graphicx}

\begin{document}

\preprint{APS/123-QED}

\title{
Generative Deep Learning for the Two-Dimensional Quantum Rotor Model.
}

\author{Yanyang Wang}
\affiliation{Key Laboratory of Quark and Lepton Physics (MOE) and Institute of Particle Physics, Central China Normal University, Wuhan 430079, China}

\author{Feng Gao}
\email[]{fgao@mails.ccnu.edu.cn}
\affiliation{Key Laboratory of Quark and Lepton Physics (MOE) and Institute of Particle Physics, Central China Normal University, Wuhan 430079, China}

\author{Kui Tuo}
\affiliation{Key Laboratory of Quark and Lepton Physics (MOE) and Institute of Particle Physics, Central China Normal University, Wuhan 430079, China}

\author{Wei Li}
\affiliation{Key Laboratory of Quark and Lepton Physics (MOE) and Institute of Particle Physics, Central China Normal University, Wuhan 430079, China}
\affiliation{ESIEA, Campus Ivry sur Seine, 73 bis Avenue Maurice Thorez, 94200
Ivry sur Seine.}

\begin{abstract}

The advancement of diverse generative deep learning models and their variants has furnished substantial insights for investigating quantum many-body problems. In this work, we design two models based on the foundational architecture of generative adversarial networks (GANs) to investigate the ground-state properties and phase transition characteristics of the two-dimensional quantum rotor model (QRM). Within a semi-supervised learning framework, we incorporate multiple layers of transposed convolutions in the generator, enabling the conditional GAN to more efficiently extract low-dimensional encoded information. Analysis of one-dimensional latent variables associated with ground-state samples for different system sizes allows us to pinpoint the location of the critical point. In addition, we introduce dynamically adaptive weighting factors related to the distributional characteristics into the loss function of the deep convolutional GAN, and utilize upsampling techniques to enlarge the generated sample sizes. Comparisons of the optimization processes for mean magnetization and potential energy density across different magnetization regimes of QRM demonstrate that our model can efficiently generate valid ground-state samples, significantly reducing computational time. Our results highlight the promising potential of generative deep learning in quantum phase transition research, especially in critical point identification and the auxiliary generation of simulation data for quantum many-body models.

\end{abstract}
\maketitle



\section{Introduction}
\label{intro}

The deep integration of machine learning and physics is fostering a new paradigm in scientific research. Through systematic parameter optimization and iterative updates, neural networks have demonstrated remarkable mapping and generalization capabilities, enabling broad applications across astronomy \cite{huerta2019enabling,0Probabilistic,smith2023astronomia,caldarola2024astrometric}, high-energy physics \cite{erbin2022characterizing,Ma2022AJT,steinheimer2019machine}, biophysics \cite{yan2020revealing,tareen2019biophysical,giulini2019deep}, complexity science \cite{maseer2023meta,xie2023simple}, and quantum physics \cite{ma2018transforming,kookani2023xpookynet,Zhang_2021}. In particular, the intrinsic probabilistic nature of quantum many-body systems aligns naturally with the data-driven foundation of deep learning \cite{stokes2023continuous}. The concept of neural-network quantum states (NQS) was introduced to represent quantum many-body wave functions using artificial neural networks \cite{carleo2017solving}. Relevant studies have demonstrated that deep NQS architectures based on fully connected networks and convolutional networks can more effectively capture the intricate features of quantum states \cite{choo2018symmetries,sharir2020deep,schmitt2020quantum,liu2019machine,levine2019quantum}. Among the broad range of machine-learning architectures, schemes based on supervised and unsupervised learning have achieved significant progress in the study of classical phase transitions \cite{Chen2022StudyOP,Shen2021SupervisedAU,shen2021machine}. In simulations of quantum many-body systems, deep learning can participate in the entire computational workflow by optimizing the ground-state wave function \cite{carleo2019machine}. Therefore, designing more schemes based on supervised and unsupervised learning is expected to further assist in identifying quantum phases and predicting quantum phase transitions.

Quantum phase transitions occur in quantum many-body systems and are driven by quantum fluctuations. A key characteristic is symmetry breaking: high-temperature phases typically exhibit higher symmetry, while low-temperature phases form ordered structures through spontaneous symmetry breaking, leading to non-thermal abrupt changes in ground state properties near absolute zero \cite{sachdev2011quantum}. Early impetus for research on quantum phase transitions primarily stemmed from experimental observations in superconductors, organic conductors, and related compounds \cite{bitko1996quantum,ruegg2008quantum,coldea2010quantum}, as well as studies on Mott metal-insulator transitions. Although significant progress has been made in real-time dynamical simulations of quantum many-body systems using quantum Monte Carlo and variational Monte Carlo methods \cite{georgescu2014quantum,gubernatis2016quantum,becca2017quantum,heyl2018dynamical}, substantial challenges remain in constructing ground states for large-scale coupled spin lattice systems. This difficulty mainly arises from the need to sample a vast number of quantum states when computing path integrals for partition functions. For instance, in coupled quantum rotor systems, even a slight increase in the number of rotors leads to an exponential growth in the required wave function sampling \cite{defenu2017criticality,jiang2022solving,medvidovic2023variational}. Consequently, finding suitable acceleration strategies and improved simulation methods has become a critical issue requiring urgent resolution.

Generative deep learning, when combined with physical prior knowledge, holds great promise in data augmentation, multiscale modeling, and the discovery of physical laws. Among them, generative adversarial networks (GANs)—comprising a generator and a discriminator—enable direct sampling from target distributions, circumventing the computational cost of traditional probabilistic inference \cite{goodfellow2020generative,creswell2018generative}. Various extensions such as deep convolutional GAN (DCGAN), wasserstein GAN, conditional GAN (CGAN), and cycle GAN have improved stability and diversity in generation \cite{jenkins2024exploring,arjovsky2017wasserstein,mirza2014conditional,9983478,pu2018removing}. In this work, we employ a CGAN to learn ground-state configurations of a two-dimensional quantum rotor model (QRM) for phase classification and critical-point identification. Furthermore, leveraging the flexibility of the DCGAN variant—with customized loss functions and upsampling techniques—we generate large-scale rotor configurations from smaller systems and validate the physical consistency of the generated samples through observable measurements.

The remainder of this paper is organized as follows. 
In Section II, we introduce the QRM and outline the fundamental features of its quantum phase transition based on the large-$N$ approximation. 
Section III describes in detail the customized variant of the Generative Adversarial Network framework and its application to the two-dimensional QRM. 
In Section IV, we present the main results, including the identification of critical points in quantum rotor systems of various sizes using a Conditional GAN and the accelerated generation of ground-state samples through a self-designed DCGAN model. 
Finally, Section V provides concluding remarks and summarizes the key findings of this work.

\section{Model}

Although fundamental quantum rotors do not exist in nature, they can effectively describe the quantum degrees of freedom of low-energy atoms. The lattice QRM is often regarded as an effective toy model for studying phase transitions in systems with a small number of electrons at low temperatures. This model characterizes the quantum dynamics of an $N$-component vector (referred to as a 'rotor') constrained by $\hat{\mathbf{n}}\cdot\hat{\mathbf{n}}=1$. The rotor can be visualized as a particle moving on an $d(d>1)$-dimensional hypersphere, and the Hamiltonian of continuous planar rotors is expressed as

\begin{equation}
H_R=\frac{J\tilde{g}}{2}\sum_i \hat{L}_i^2 - J\sum_{\langle i j\rangle}\hat{\mathbf{n}}_i\cdot\hat{\mathbf{n}}_j.
\end{equation}

Here, $\hat{\mathbf{n}}=(\cos\theta_i,\sin\theta_i)$ is a unit vector operator, and $L_i=-i\partial_{\theta_i}$ denotes the invariant operator formed by the angular momentum tensor in the asymmetric rotor space. $J$ represents the coupling between neighboring rotor orientations, and $\tilde{g}$ is a constant parameter. When the rotor orientations are highly uncertain, the kinetic term (the first term of $H_R$) dominates, delocalizing individual rotors and driving the system toward a magnetically disordered state. Conversely, the nearest-neighbor coupling term favors ferromagnetic alignment, leading to an ordered phase. Therefore, the control parameter $g$ determines the quantum phase characteristics of the system. When $g$ is sufficiently small such that the first term in $H_R$ becomes negligible, the equilibrium state exhibits magnetic order. In the continuous basis representation, the evolution of the quantum state $\Psi=\psi(\theta)$ is governed by

\begin{equation}
i\frac{\partial\Psi}{\partial t}=-\frac{gJ}{2}\sum_k \frac{\partial^2\Psi}{\partial\theta_k^2} - J\sum_{\langle k,l\rangle}\cos(\theta_k-\theta_l)\Psi.
\label{e_2}
\end{equation}

Even for a small number of interacting rotors, Eq.~(\ref{e_2}) is difficult to solve exactly, and the continuous nature of the basis further increases this complexity. Consequently, various analytical approximations and numerical simulations have been developed to study the QRM.

The large-$N$ expansion, originally developed in the context of classical models \cite{brezin1973critical}, is a non-perturbative approximation technique that has found wide application in quantum field theory and statistical mechanics. By extending the intrinsic symmetry group of a system from $SU(2)$ or $SU(3)$ to $SU(N)$ and taking the limit of large $N$, the original theory becomes analytically tractable. This method is particularly valuable for studying quantum phase transitions, which occur near absolute zero and are driven by quantum rather than thermal fluctuations, making them difficult to treat using standard perturbative methods. For the QRM, the large-$N$ approximation provides a reliable description of the phase diagram and static observables at zero temperature, offering quantitative insight into quantum criticality.  

Under the assumption that the order parameter is polarized along the basis direction, the path-integral form of the partition function is given by
\begin{equation}
\begin{aligned}
\mathcal{Z} = \int \mathcal{D}\lambda\,\mathcal{D}r_0 \exp \Big[ 
& -\frac{N-1}{2}\operatorname{Tr}\ln(-c^2\partial_i^2 - \partial_\tau^2 + i\lambda) \\
& + \frac{iN}{cg}\int_0^{1/T} d\tau \int d^d x\, \lambda(1-r_0^2) \Big].
\end{aligned}
\end{equation}

Here, the Lagrange multiplier $\lambda$ enforces the constraint $n^2=1$. For $T=0$ and $d>1$, evaluating the integral of the partition function leads to the saddle-point equations

\begin{equation}
\begin{aligned}
N_0^2 + g \int^{\Lambda}\frac{d^{d+1}p}{(2\pi)^{d+1}}\frac{1}{p^2+(m/c)^2} &= 1, \\
m^2 N_0 &= 0,
\end{aligned}
\end{equation}

where $i\lambda=m^2$ defines the Lagrange multiplier. Analyzing the analytical solution of these equations reveals the existence of a critical coupling $g_c$ in the quantum rotor system. As $m\rightarrow0$, the integral increases monotonically with decreasing $m$, and the equation admits a solution when the integral reaches its maximum finite value:
\begin{equation}
\begin{aligned}
m &= 0, \\
N_0^2 &= 1 - g\int^{\Lambda}\frac{d^{d+1}p}{(2\pi)^{d+1}}\frac{1}{p^2} \\
&= 1 - \frac{g}{g_c}.
\end{aligned}
\end{equation}

The quantum critical scaling of the order parameter is then expressed as $N_0 \sim (g_c - g)^{\beta}$, which decays algebraically as the system approaches the critical point. Similarly, the spin stiffness $\rho_s$ (also known as the helicity modulus), representing the system’s resistance to spatial spin distortions, exhibits the scaling behavior $\rho_s \sim (g_c - g)^{(d-1)\nu}$. These results indicate that near the quantum critical point, the spin orientations undergo large-scale rearrangements corresponding to transitions between quantum paramagnetic and ferromagnetic phases. Therefore, for the two-dimensional QRM, a quantum phase transition occurs as the control parameter $g$ is varied, and distinct quantum phases can stably exist away from the critical region. This theoretical framework provides essential support for applying a semi-supervised GAN-based approach to identify quantum phases and locate critical points in our study.

\section{Method}


\begin{figure*}[t]
    \centering
        \includegraphics[width=0.8\textwidth]{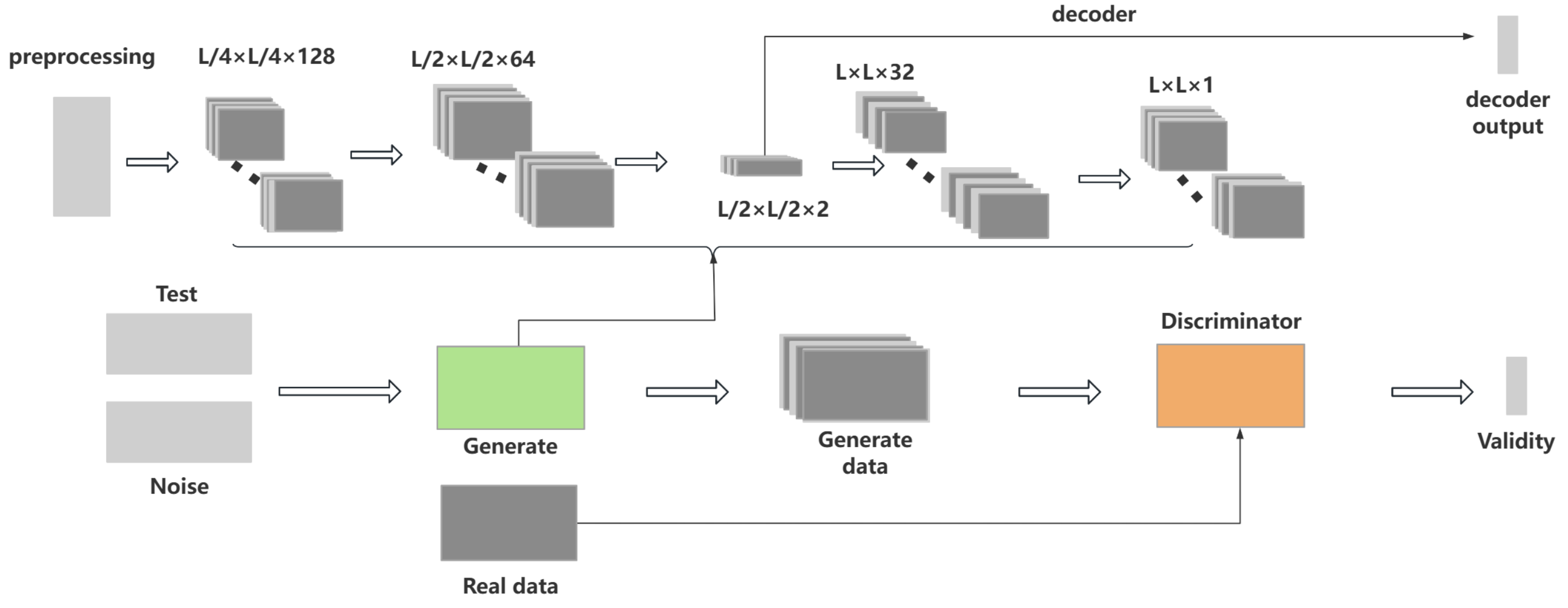} 
\caption{Schematic diagram of the conditional generative adversarial network structure with extractable generator-encoding information. The lower bracket illustrates the general CGAN training workflow, while the upper bracket highlights the details of the transposed convolutional operations added to the generator. The input data are preprocessed and successively passed through convolutional layers for dimensionality reduction and transposed convolutional layers for upsampling, with corresponding adaptations made to the discriminator for compatibility. After training, both two-dimensional and one-dimensional encoded representations are extracted from the generator for analysis.}

\label{f_1}
\end{figure*}

In this section, we describe the overall procedure for obtaining ground-state samples of the QRM and introduce the two generative deep-learning frameworks developed in this work. The first approach, based on the concept of semi-supervised learning, constructs a complete CGAN architecture to identify the critical points of the two-dimensional QRM across different system sizes. The second approach, formulated within an unsupervised learning framework, employs a customized DCGAN with a self-defined loss function to accelerate the generation of large-scale ground-state samples. The design principles and application rationale of these two GAN variants are presented in the method section, while the results section further elaborates on the concrete details of the study.

Reference \cite{medvidovic2023variational} proposed a convolutional neural network (CNN)-based framework for simulating the real-time dynamics of quantum rotor systems, representing a typical application of NQS. The central idea of NQS is to represent the wavefunction of a quantum many-body system using an artificial neural network characterized by a set of internal parameters $w$, providing an optimal representation of the ground state and its time-dependent observables under a given Hamiltonian $H$. Once the network is trained, the resulting machine-learning architecture can output both the phase and amplitude of the wavefunction $\Psi$ for a given quantum configuration.  

The optimization of neural network parameters is grounded in the variational principle, where a feedback mechanism is employed to approximate both the time-evolving and ground-state wavefunctions. The variational monte carlo (VMC) simulation method \cite{misawa2019mvmc} is utilized in this optimization process, performing stochastic sampling over the network parameters to minimize the energy expectation value:
\begin{equation}
E(w) = \frac{\left\langle \Psi_\alpha \right| \mathcal{H} \left| \Psi_\alpha \right\rangle}
{\left\langle \Psi_\alpha | \Psi_\alpha \right\rangle}.
\end{equation}
Through iterative sampling, the energy gradient can be estimated, allowing parameter updates via gradient descent and achieving a self-consistent optimization of the wavefunction.

For the QRM, we employ a CNN architecture in which the sine and cosine values of the relative angular displacements of the rotors are used as input features. The CNN encodes these inputs and reduces them to a single complex-valued output representing the natural logarithm of the wavefunction, $\ln \Psi_\alpha(\theta)$. However, the stochastic parameter updates required for optimizing wavefunctions in large-scale rotor systems become computationally expensive. Because stochastic updates of network parameters become prohibitively time-consuming when optimizing large-scale rotor-wave functions, incorporating hamiltonian monte carlo (HMC) is necessary to accelerate the optimization. The basic idea of HMC is to introduce a sampling path in the parameter space, enhancing sampling efficiency by using the gradient information of the target distribution to guide proposal updates, thereby biasing sampling toward regions of higher probability density \cite{betancourt2017conceptual}. 

By initializing the QRM with a control parameter $g$ and generating a random representation of the corresponding wavefunction from its hamiltonian, the CNN architecture iteratively updates the network parameters through the VMC process, while HMC directs the parameter updates. This combined approach yields ground-state samples of the QRM for different lattice sizes and coupling parameters $g$, where each sample encodes the angular value of every rotor. From these samples, observable quantities such as the average ground-state magnetization $M$ can be evaluated. The resulting ground-state configurations under various $g$ values constitute the training dataset, which is subsequently used as training data for the variant GANs described in the following sections.

The deep-learning framework employed in this work is primarily based on variant generative models derived from the GAN architecture. The original GAN was proposed to address the intractable probabilistic computations encountered in deep generative models under maximum-likelihood estimation and related strategies. Its adversarial mechanism eliminates the need for explicit feedback loops during generation, thereby improving the efficiency of gradient backpropagation through the use of piecewise linear activation functions. The training objective of a GAN is defined by the following minimax optimization function:

\begin{equation}
\begin{aligned}
\min _G \max _D V(D, G)=\mathbb{E}_{\boldsymbol{x} \sim p_{\mathrm{data}}(\boldsymbol{x})}[\log D(\boldsymbol{x})]+ \\
\mathbb{E}_{\boldsymbol{z} \sim p_{\boldsymbol{z}}(\boldsymbol{z})}[\log (1-D(G(\boldsymbol{z})))] .
\label{equ2}
\end{aligned}
\end{equation}

Here, $G(\boldsymbol{z})$ represents the mapping of a latent prior distribution $p(\boldsymbol{z})$ to the data space, and $D(\boldsymbol{x})$ denotes the probability that a sample $\boldsymbol{x}$ originates from the true data distribution. Both $G$ and $D$ are nonlinear mapping functions, allowing flexibility in designing the generator and discriminator architectures. The optimization process seeks to simultaneously maximize $D$ and minimize $\log(1 - D(G(\boldsymbol{z})))$. Specifically, the generator aims to produce samples that are indistinguishable from real data, while the discriminator strives to correctly distinguish real from generated samples and satisfy given conditions. During training, these two objectives are alternately optimized through a min–max process.  

In our study, the Conditional GAN model is employed for identifying the critical points of the QRM, while a customized Deep Convolutional GAN architecture is developed to accelerate the generation of large-scale equilibrium-state samples. Both approaches share the same fundamental GAN structure and thus exhibit similar training dynamics.  

Among various GAN variants, the CGAN introduces additional conditional factors during training to enhance the interpretability and controllability of the generated results. The conditional information is closely coupled with the generated distribution: upon receiving the condition input, the generator incorporates it into the generated samples through its internal neural network structure, ensuring that the outputs not only resemble real data but also reflect the prescribed conditions. In our implementation, we employ conditional batch normalization, where the conditional information modulates the normalization parameters at each layer of the generator. Based on theoretical insights from the QRM, we assign partial labels to the training dataset in regions far from the quantum critical regime, thereby improving the CGAN’s capability to distinguish between different quantum phases.

The large-$N$ approximation confirms the critical characteristics of the two-dimensional quantum rotor system, demonstrating that the quantum paramagnetic and ferromagnetic phases remain stable in the non-critical regime away from the control parameter $g_c$. Based on this theoretical insight, we employ a semi-supervised learning approach to label equilibrium-state samples corresponding to different coupling parameters $g$, and construct a complete training dataset for input into the CGAN. The customized generator and discriminator architectures are illustrated in Fig.~\ref{f_1}. The generator performs convolutional encoding to reduce the dimensionality of the input data, followed by transposed convolution to reconstruct the output to its original spatial dimensions. The discriminator, on the other hand, applies downsampling convolutions to process and reduce the dimensionality of the generated images. After sufficient training epochs, we extract the generator’s encoded representations (either in one or two dimensions) and the discriminator’s validity outputs for both real and generated samples. By statistically averaging these network outputs across different values of $g$, we can identify the emergence of the critical region and determine the precise location of the critical point.

The Deep Convolutional Generative Adversarial Network is another important variant of the GAN framework, originally proposed to address several issues observed in early GANs, including instability during training, limited quality of generated samples, and poor model controllability. The core idea of DCGANs is to leverage the powerful feature extraction and hierarchical spatial learning capabilities of convolutional neural networks to construct more stable and efficient adversarial models. In our implementation, the customized DCGAN architecture closely follows the generator and discriminator design of the CGAN model shown in Fig.~\ref{f_1}. Since our goal is to accelerate the generation of large-scale ground-state samples of the QRM within an unsupervised learning framework, several key differences distinguish the DCGAN from the CGAN in Fig.~\ref{f_1}.  

First, all label information is removed when constructing the training dataset. Both the training and testing datasets consist solely of raw ground-state samples corresponding to different coupling parameters $g$, and the generator and discriminator are adaptively adjusted to accommodate this unlabeled configuration. Second, instead of indexing the encoded latent features in the generator, we employ transposed convolutions to upsample the generator’s output to a target spatial size of $2L \times 2L$, corresponding to the output dimension of the generator in the upper panel of Fig.~\ref{f_1}. Finally, we design a customized loss function for the DCGAN by extending the standard binary cross-entropy loss with a dynamic weighting term that captures the central tendency, dispersion, and shape characteristics of the generated data distribution.  

The primary objective of constructing this customized DCGAN architecture is to efficiently generate large-scale ($2L$) ground-state samples from smaller-size ($L$) quantum rotor configurations. This approach significantly reduces the computational cost associated with optimization as the number of rotors increases, providing a practical and scalable solution for modeling high-dimensional quantum systems.

\section{Result}
\subsection{Determination of the critical point}

\begin{figure}
\centering
\includegraphics[width=0.45\textwidth]{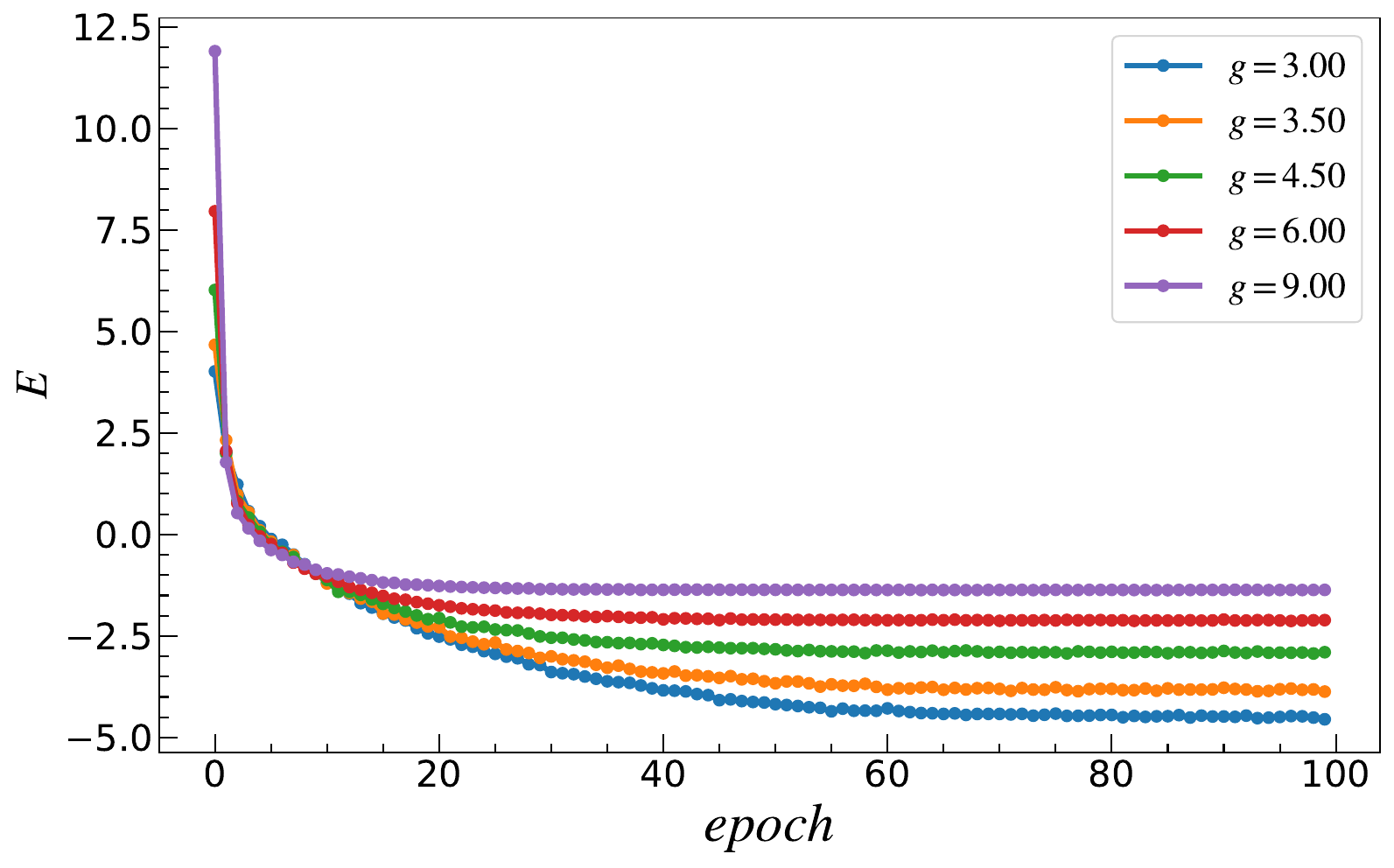}
\caption{
Optimization process of the ground-state energy for the two-dimensional QRM with $L = 4$. For different coupling parameters $g$, the sample energies gradually converge and stabilize after sufficient optimization steps, indicating that the system reaches its ground state. The optimized parameters corresponding to the ground state are then stored and used to generate a large number of samples for constructing the training dataset.
}

\label{f_2}
\end{figure}

\begin{figure}
\centering
\includegraphics[width=0.45\textwidth]{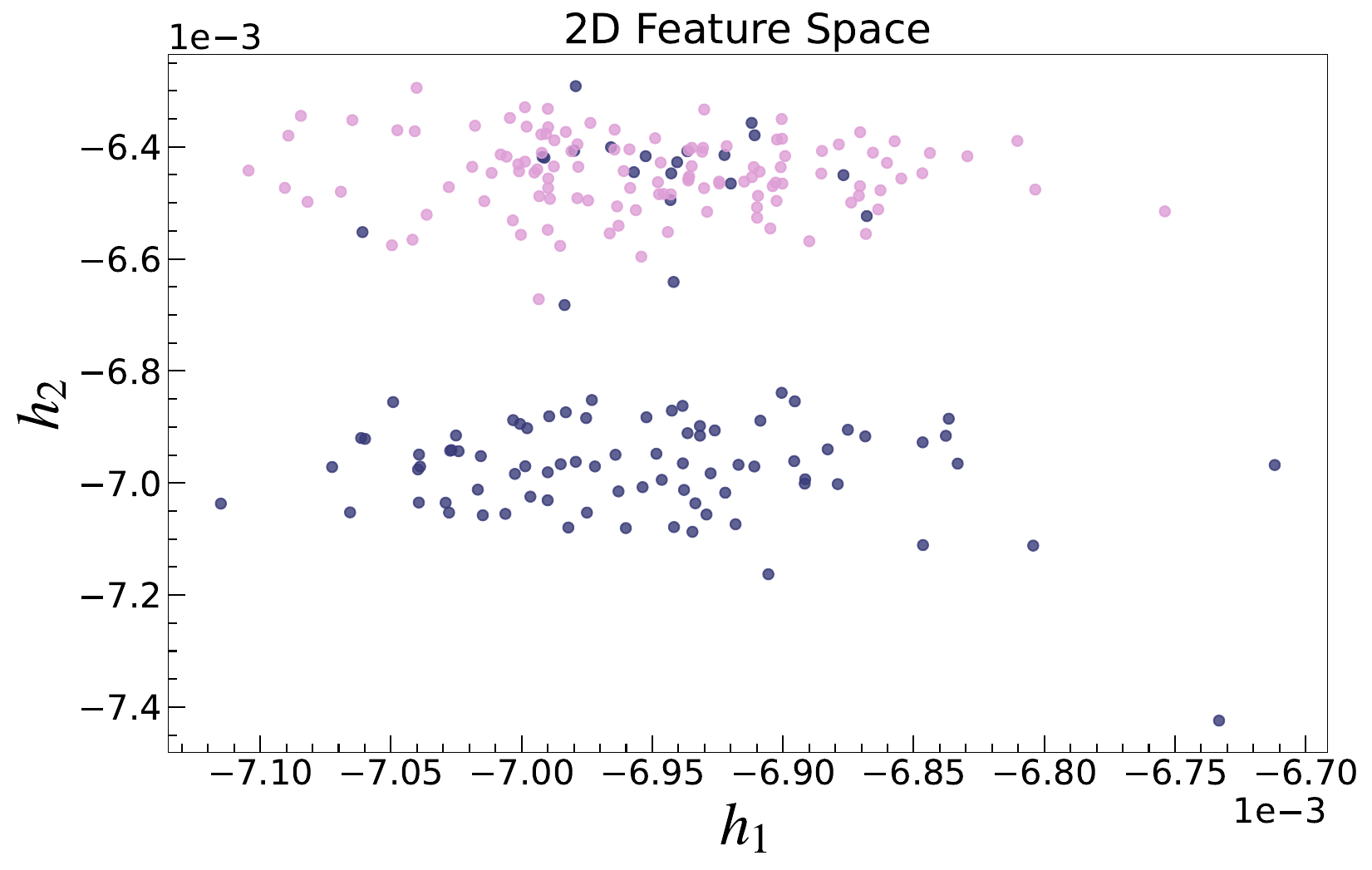}
\caption{
After CGAN training on a series of ground-state samples of the two-dimensional QRM at various $g$ values, the generator’s 2-D encoded representations for the test set were extracted. Different colored points represent the discriminator’s predicted classes. Based on these classifications, the generator has learned to extract statistics that distinguish between the quantum paramagnetic phase and the quantum diamagnetic phase.
}

\label{f_3}
\end{figure}

Using the dynamic optimization framework of NQS, we obtained the ground-state energies of the QRM for small system sizes, which serve as training samples for the subsequent neural network analysis. The optimized ground-state energies for different coupling parameters $g$ are shown in Fig.~\ref{f_2}. The optimization process begins by defining a variational wavefunction and initializing its associated parameters. For a given Hamiltonian and simulation size, the gradients are computed jointly with the variational parameters to minimize the energy expectation value. During the optimization, the HMC method is employed to determine the update direction and simultaneously perform efficient sampling of the quantum rotor configurations.  

The dataset consists of 10,000 matrices of size $4 \times 4$, where each matrix element represents the relative angular displacement of an individual quantum rotor, determined within a fixed Hilbert-space basis. As illustrated in Fig.~\ref{f_2}, the ground-state energy gradually converges and stabilizes after a sufficient number of optimization steps. To ensure numerical stability and consistency, the final set of optimized parameters is used to generate a large number of ground-state samples, which are subsequently employed to construct the training and testing datasets for the GAN-based models.

\begin{figure*}[t]
\begin{tabular}{cc}
    \includegraphics[width=0.45\textwidth]{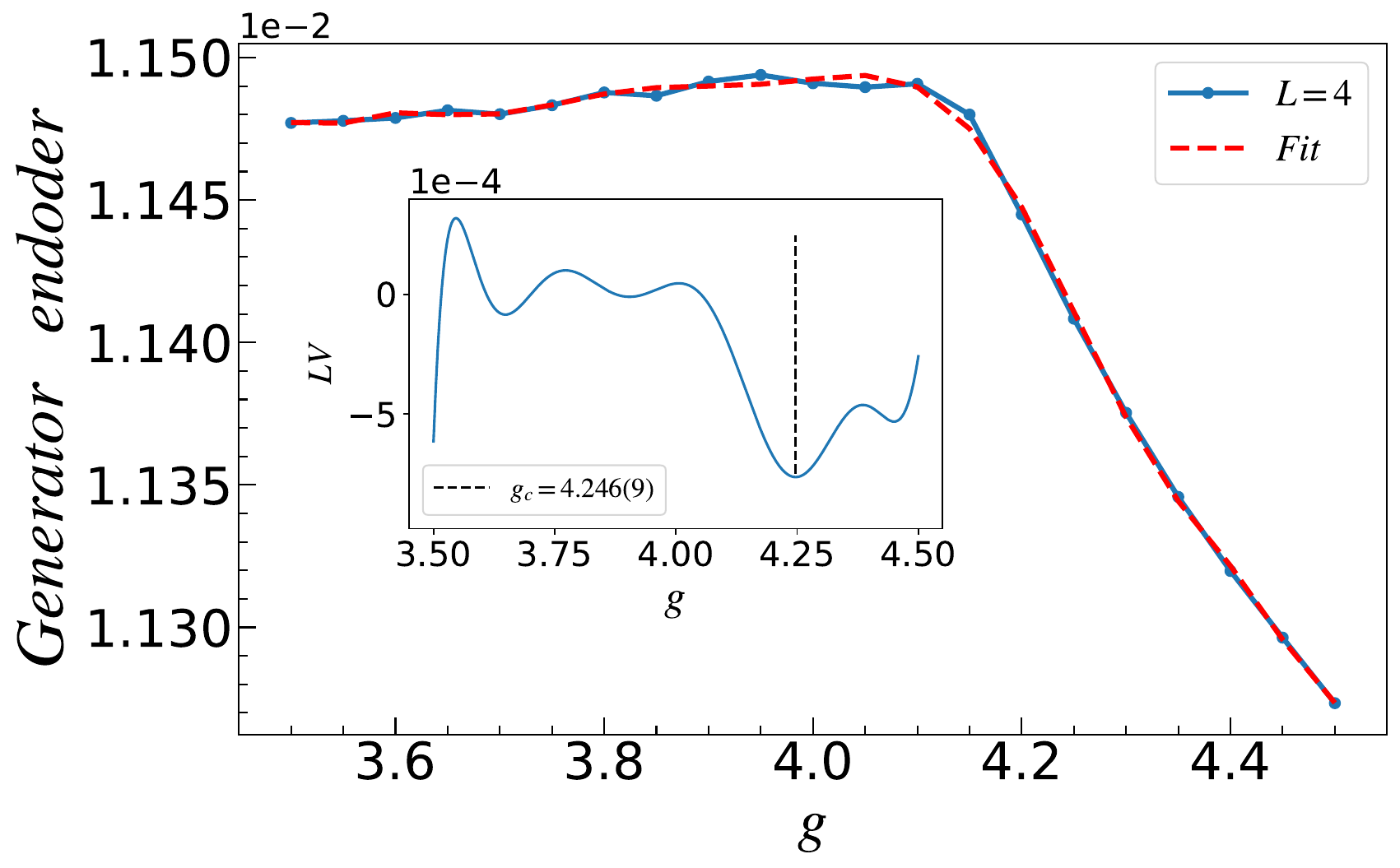} &
    $\qquad$\includegraphics[width=0.48\textwidth]{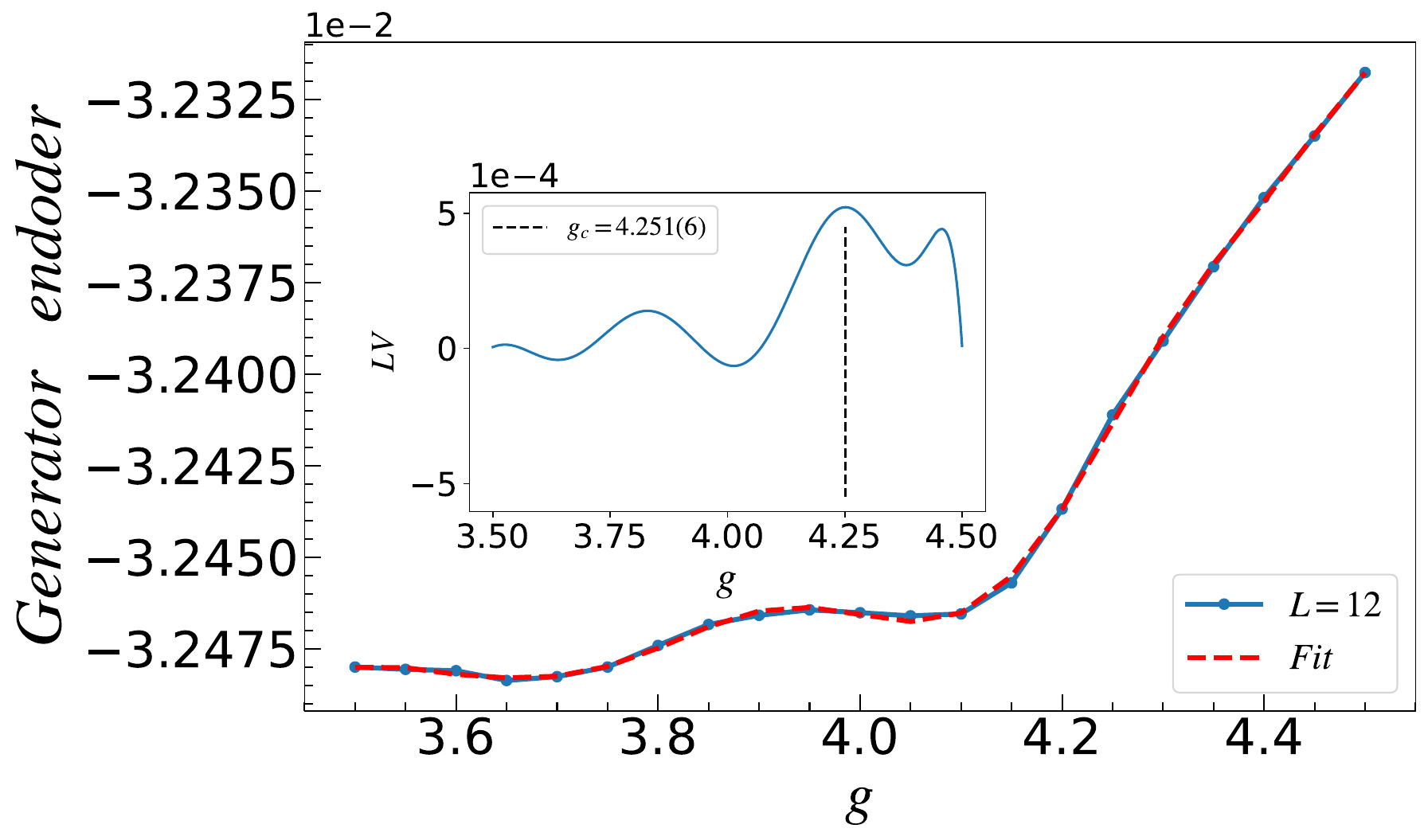} \\
    {\quad}(a) & $\qquad$ {\quad}(b)
\end{tabular}
\caption{
(a) One-dimensional encoded information extracted from the generator for the system size $L \times L = 4 \times 4$, obtained from the test set composed of ground-state samples at various coupling parameters $g$. The red dashed line represents the polynomial fitting curve, while the blue line in the inset shows the curvature of the fitted curve. By analyzing the hidden variable $LV$, we find that the minimum curvature extremum effectively characterizes the location of the critical point, consistent with the physical behavior of the QRM, where spin orientations undergo drastic changes near the critical region. (b) One-dimensional encoded information from the generator for $L \times L = 12 \times 12$. Unlike Fig.~\ref{f_4}(a), the fitted curve exhibits opposite monotonicity, and the maximum curvature extremum corresponds to the point of the most rapid change in the fitted curve, allowing the identification of the critical coupling $g_c$.}
\label{f_4}
\end{figure*}

For the semi-supervised learning framework, it is necessary to incorporate prior knowledge of the QRM, informed by both theoretical analysis and partial numerical simulations. The construction of the training dataset under a series of coupling parameters $g$ proceeds as follows. We first define the range of $g$ values within the interval $g \in [3.5, 4.5]$, with increments of $\Delta g = 0.05$. After a sufficient number of optimization steps, the ground-state parameters are used to generate 1,000 datasets of shape $(10000, L, L)$ as training samples, together with 100 ground-state samples under the same parameter settings as the testing dataset.  

Partial labeling is then applied to the training data. Specifically, the ground-state samples corresponding to $g \in [3.50, 3.60]$ are labeled as ‘0’, while those corresponding to $g \in [4.40, 4.50]$ are labeled as ‘1’. This labeling strategy distinguishes the features at the two extremes of the control-parameter range within a narrow interval, reflecting the contrasting magnetic characteristics of the QRM in different coupling regimes. Such partial supervision enables the CGAN to learn phase-discriminative representations while maintaining generalization capability in the unlabeled intermediate region near the critical point.

After constructing the training dataset for the CGAN, random noise vectors are fed into the generator, where they are processed through convolutional layers and reduced to one- or two-dimensional latent representations that are indexed internally. The generator then applies transposed convolutions to reconstruct these latent features into output samples. Real samples with partial labels are simultaneously introduced into the CGAN training workflow, serving to optimize the generator’s output quality and assist the discriminator in making more accurate decisions. The discriminator merges the generated samples with the corresponding label features at appropriate dimensional levels, followed by convolutional and fully connected layers to output a validity score indicating whether the input is real or generated.  

The network is trained over a sufficient number of epochs until both accuracy and loss converge and remain stable across extended training. During the CGAN construction and training process, appropriate choices of the loss function and optimizer are critical. Given the classification nature of our phase-identification task, the binary cross-entropy loss is adopted as an effective criterion to jointly account for the influence of both labeled and unlabeled samples. To prevent vanishing or exploding gradients and ensure stable convergence, the Adam optimizer is employed, which adaptively adjusts the learning rate based on the first and second moments of the gradients. This adaptive learning mechanism considerably accelerates training, particularly for large-scale datasets of quantum rotor ground-state samples.

The trained CGAN model is then applied to the testing dataset to extract the encoded representations from the generator. The two-dimensional encoded results for $L=4$ are shown in Fig.~\ref{f_3}, where different colors correspond to the CGAN’s classification predictions for the test samples. As illustrated in the figure, even with only partial labeling, the trained CGAN successfully classifies unlabeled equilibrium-state samples, effectively distinguishing the quantum ferromagnetic and paramagnetic phases of the rotor model. Although a few data points in the quantum critical region remain difficult to categorize precisely, the generator, through its adversarial optimization with the discriminator, is capable of extracting statistical features that characterize the distinct quantum phases. Consequently,we analyze the one-dimensional encoded data obtained from the generator and use the variations of the encoded representations with respect to different $g$ values to identify the critical coupling point.

Based on the trained CGAN model, the encoded representations obtained from the generator successfully classify the equilibrium-state samples of the QRM. The regions where the two-dimensional data points exhibit classification ambiguity contain crucial information about quantum criticality. Our strategy for determining the critical coupling $g_c$ from the one-dimensional encoded data of the generator proceeds as follows. We first compute the statistical average of the one-dimensional encoded outputs obtained from up to 100 test matrices of size $(1000, 4, 4)$. The averaged results are shown as blue dots in Fig.~\ref{f_4}(a). To capture the variation of the encoded data, a polynomial fitting is applied, and the red dashed line in Fig.~\ref{f_4}(a) represents the fitted curve. The inset in Fig.~\ref{f_4}(a) displays the curvature of this fitted function. By analyzing the curvature extrema of the fitted curves across different system sizes, we find that the maximum or minimum curvature values can serve as robust indicators of the system’s critical point. Figure~\ref{f_4}(b) illustrates this analysis for the two-dimensional rotor system with $L=12$, showing both the encoded data and the curvature-derived identification of the critical region.  

For the QRM, the spin stiffness $\rho_s$ exhibits characteristic scaling behavior at the quantum critical point, where the spin orientations undergo significant large-scale rearrangements. In finite-size numerical simulations, the magnetic susceptibility also shows pronounced variation relative to system size. The CGAN model captures these sensitive features, which are reflected in the generator’s encoded representations. Depending on the monotonicity of the fitted function, either the maximum or minimum of the curvature corresponds to the point of most rapid change, directly associated with the onset of the quantum phase transition. As shown in Fig.~\ref{f_4}, the curvature extrema of the one-dimensional encoded data indicate that for the $L=4$ system, the critical coupling is located at $g_c = 4.246(9)$, while for $L=12$, the critical point is $g_c = 4.251(6)$.  

\begin{figure}
\centering
\includegraphics[width=0.45\textwidth]{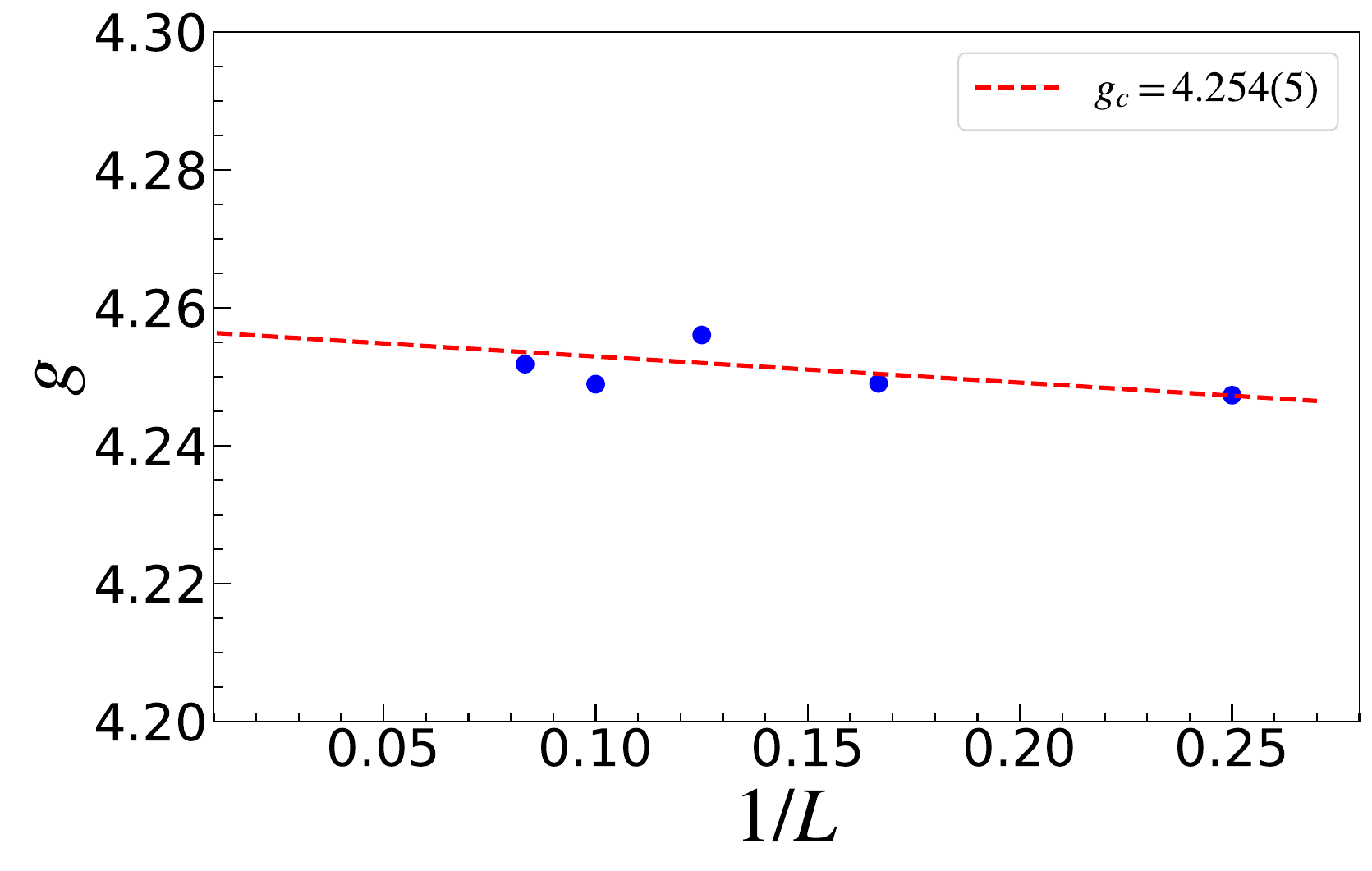}
\caption{
Determination of the critical point and linear regression for system sizes $L = 4, 6, 8, 10, 12$. Based on the finite-size data of the two‑dimensional QRM, the y-intercept of the red line obtained from the fit provides a reasonable extrapolation of the critical point to the thermodynamic (infinite-size) limit.}
\label{f_5}
\end{figure}

To minimize finite-size effects in determining the quantum critical point, we evaluated $g_c$ for system sizes $L = 4, 6, 8, 10,$ and $12$, as summarized in Fig.~\ref{f_5}. A linear regression extrapolation to the infinite-size limit yields an estimated critical coupling of $g_c = 4.254(5)$ for the two-dimensional QRM. This consistency across system sizes demonstrates that the CGAN-encoded representation effectively captures quantum critical behavior and provides a reliable data-driven means to identify the critical point.

\subsection{Accelerated generation of large-scale rotor coupled systems}

\begin{figure*}[t]
\begin{tabular}{cc}
    \includegraphics[width=0.47\textwidth]{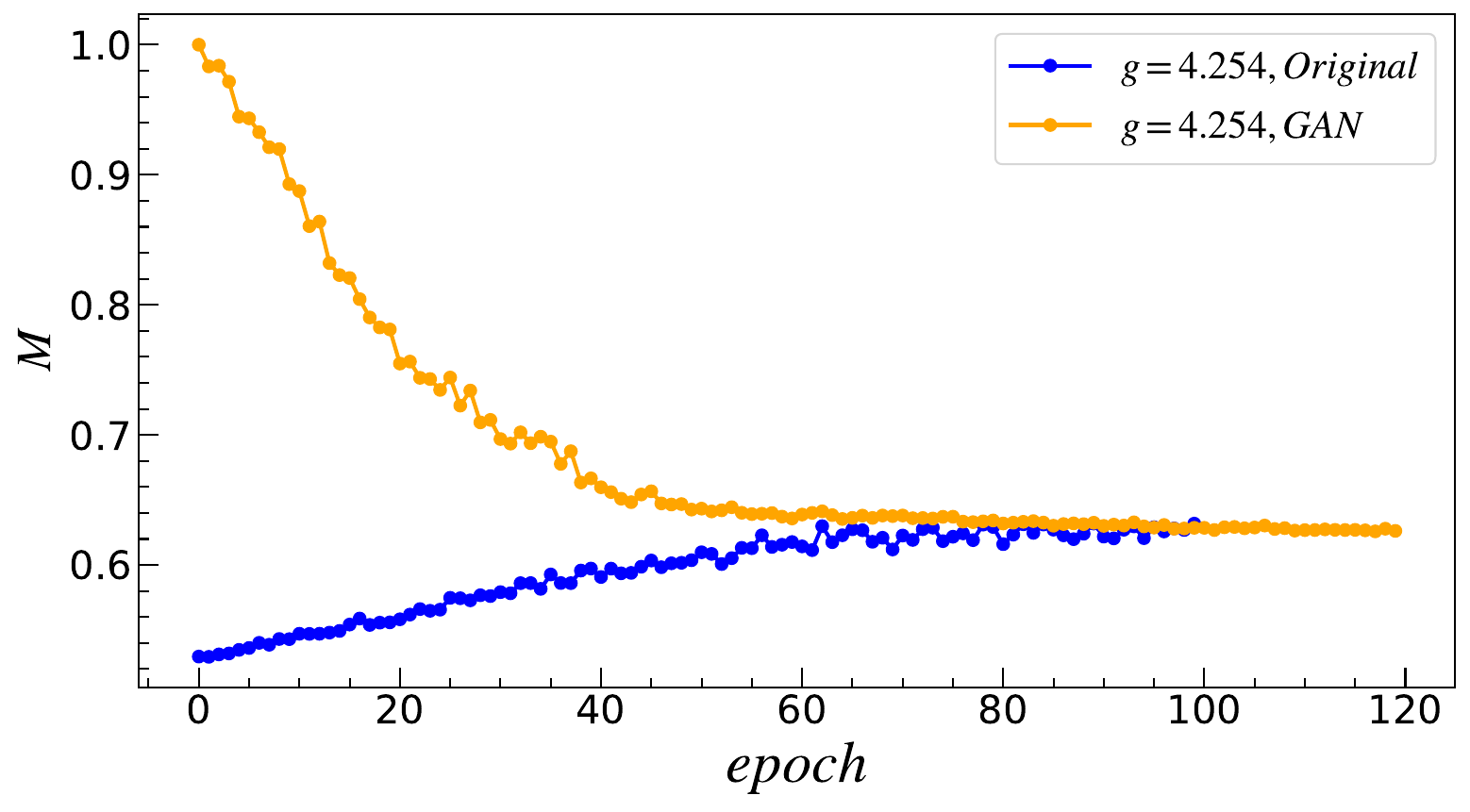} &
    $\qquad$\includegraphics[width=0.46\textwidth]{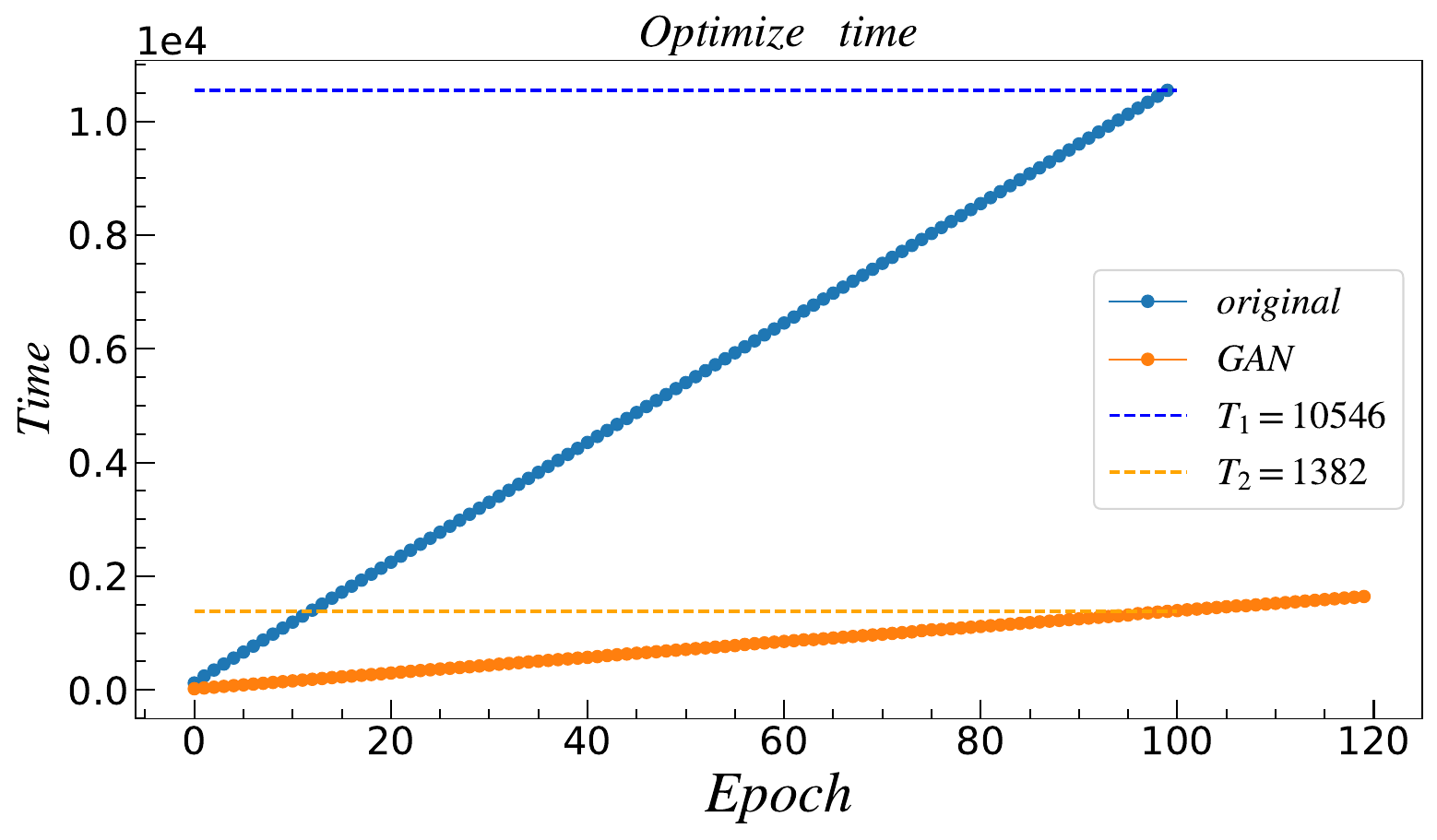} \\
    {\quad}(a) & $\qquad$ {\quad}(b)
\end{tabular}
\caption{
(a) Accelerating the generation and measurement of the mean magnetization of ground-state samples for the two-dimensional QRM using a self-designed DCGAN architecture. In the figure, the orange curve denotes the system-averaged magnetization obtained from samples generated by the DCGAN model, while the blue curve denotes the magnetization of samples obtained by optimization with the neural-network quantum state (NNQS) architecture for the same system size. It can be seen that, with sufficiently optimized training epochs, the DCGAN can produce ground-state samples that are consistent with the measured observables. (b) A time-cost comparison for obtaining ground-state samples at system size $L=8$ in Fig. 6(a), where the vertical axis is in seconds.}
\label{f_6}
\end{figure*}

\begin{figure*}[t]
\begin{tabular}{cc}
    \includegraphics[width=0.46\textwidth]{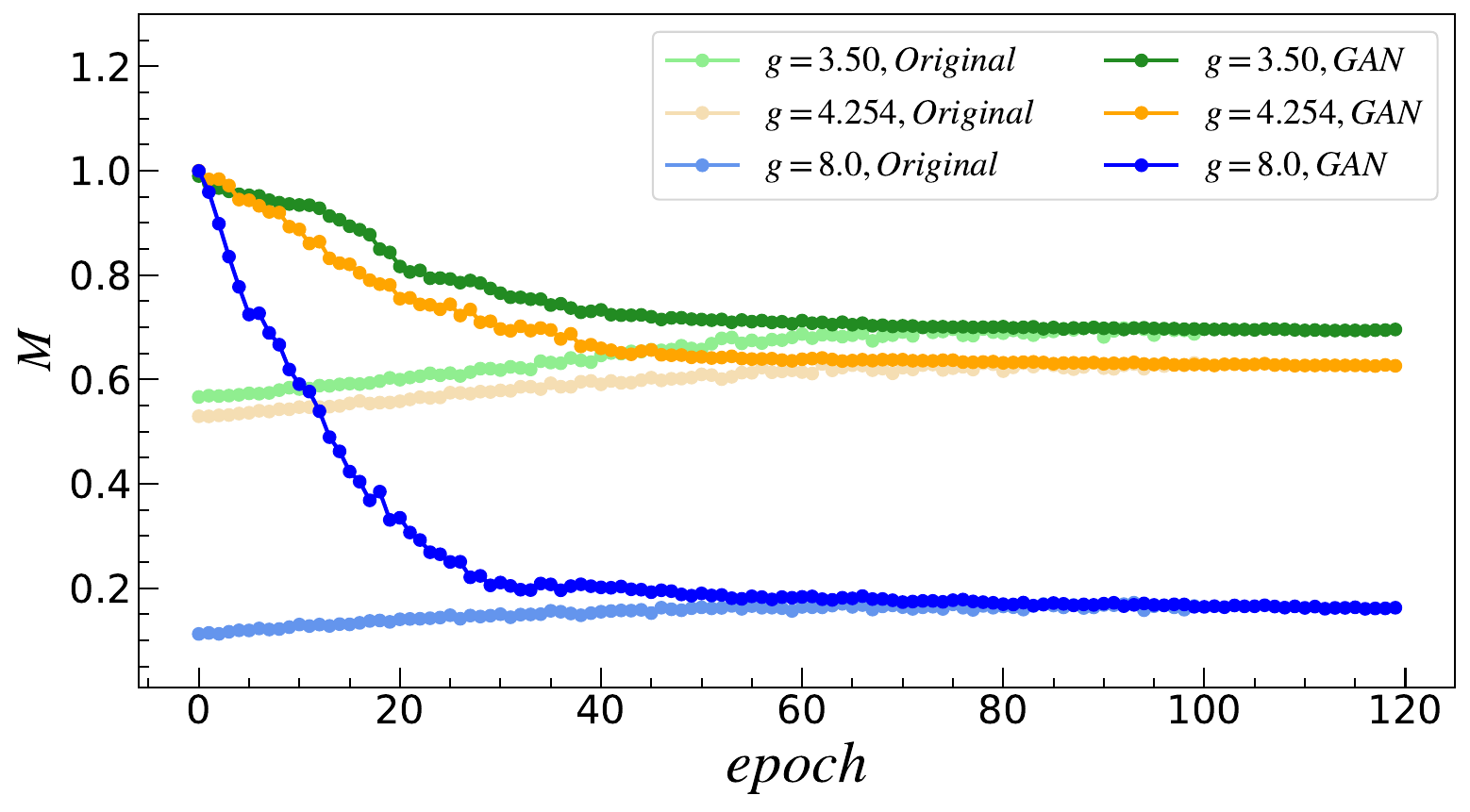} &
    $\qquad$\includegraphics[width=0.48\textwidth]{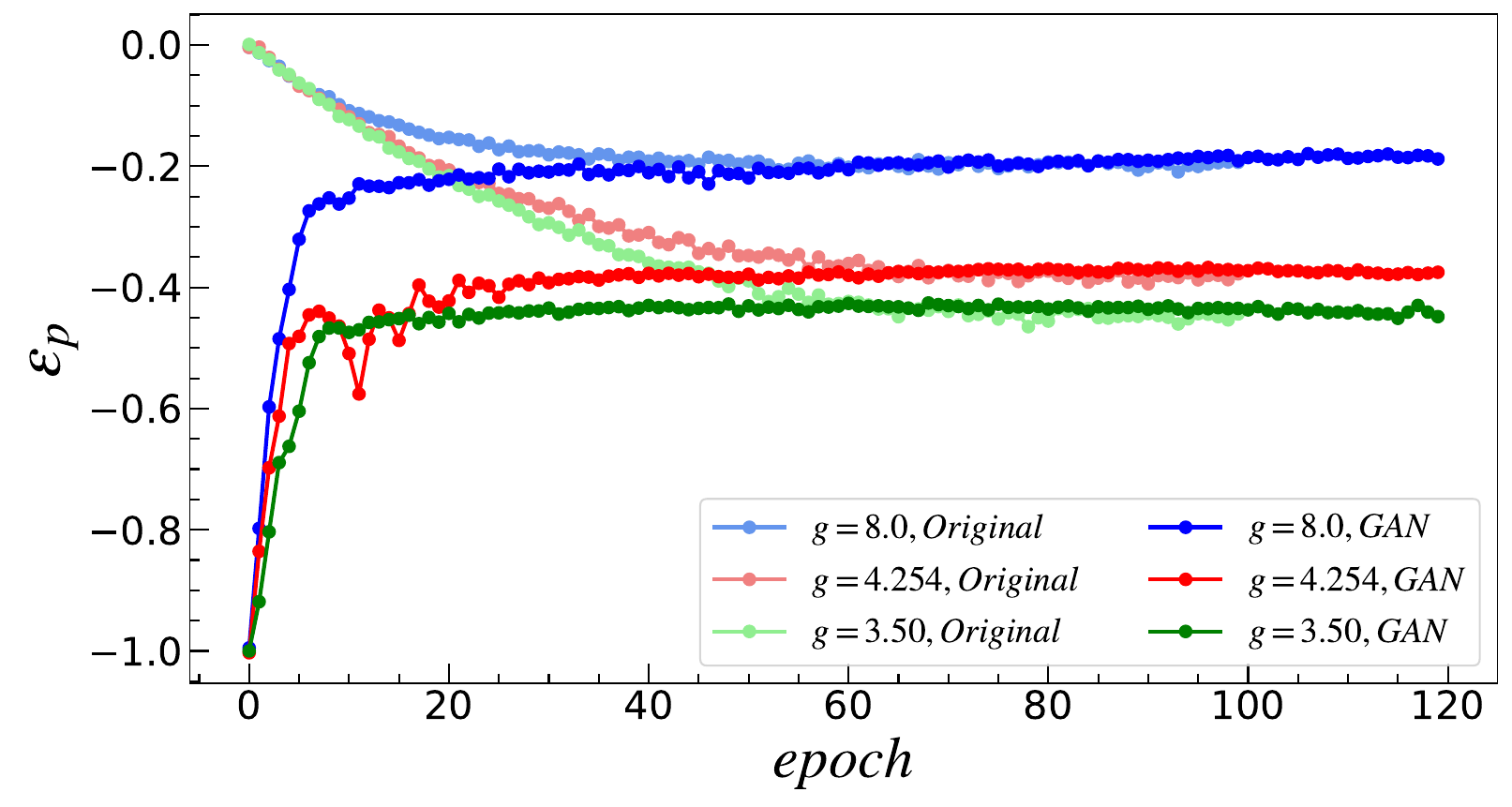} \\
    {\quad}(a) & $\qquad$ {\quad}(b)
\end{tabular}
\caption{
(a) Comparison of the average magnetization obtained from ground-state samples of the two-dimensional QRM under different magnetic states. The lighter ascending curves correspond to the variation of the average magnetization obtained using the original optimization method, while the darker descending curves represent the results generated by the DCGAN model. For systems in different magnetic states corresponding to various $g$ values, the DCGAN reproduces ground-state samples that are consistent with the original data at the level of observable quantities. (b) Measurement of the potential energy density during the DCGAN optimization process. The lighter and darker curves denote the results from the original method and the DCGAN, respectively. After sufficient optimization steps, the consistent evolution of potential energy density at the same statistical level further confirms the validity of the DCGAN-generated ground-state samples.}
\label{f_7}
\end{figure*}

The primary objective of this work is to design a generative deep-learning framework based on the DCGAN architecture to generate large-scale ground-state samples of the QRM that share the same statistical and distributional properties as the original small-size datasets. The main innovations of our customized DCGAN training process include the implementation of an upsampling technique and the design of a modified loss function.

Within the standard GAN framework, the procedure for generating large-scale ground-state samples using our customized DCGAN is as follows. The equilibrium-state samples of the two-dimensional QRM with a small lattice size $L \times L$ are used as the training dataset. Unlike the CGAN architecture, this dataset contains no label information and is directly fed into the discriminator as real data. In the generator, random noise vectors are passed through multiple transposed convolution layers, where an upsampling operation is introduced to increase the output dimension from $L \times L$ to $2L \times 2L$.  

To construct the complete DCGAN architecture, we define a customized pixel-statistics-based loss function that extends the conventional binary cross-entropy (BCE) loss by introducing additional weighting terms describing the central tendency, dispersion, and skewness of the generated data distribution. The loss function is expressed as

\begin{equation}
L = w_1[BCE] + w_2\mu + w_3\sigma + w_4[Sk],
\label{f_8}
\end{equation}

where $BCE$ denotes the binary cross-entropy, $\mu$ represents the mean of the generated data distribution, $\sigma$ is the standard deviation, and $Sk$ denotes the skewness. The coefficients $w_i$ serve as dynamic weighting factors that are adaptively updated during the DCGAN training process. After completing the training, we extract the generated output samples from the generator and record the corresponding training time for performance evaluation.

We focus on the following adjustments in DCGAN to ensure training stability and robustness of the generated results. Batch normalization is applied to the hidden layers of both the generator and the discriminator. Batch normalization stabilizes the training process, accelerates convergence, and mitigates the problem of gradient explosion. Normalizing the activations of each layer helps keep the inputs to each layer on a comparable scale. LeakyReLU activations are used in all layers except the output layer. LeakyReLU (Leaky Rectified Linear Unit) is a variant of ReLU that has a small nonzero slope in the negative region, thereby addressing the zero-gradient issue for negative inputs and allowing a small amount of negative gradient to pass, which helps prevent neurons from dying. In terms of information-preserving design for the generator and discriminator, we employ fewer pooling layers and more convolutional layers to retain more spatial information, thereby learning finer features of the real data distribution. Since the GAN training objective is the loss function (\ref{f_8}), whose global optimum is the complete replication of the real data generation process, resulting in generated data that are indistinguishable from real data, this customized DCGAN training procedure can learn the detailed ground-state sample distribution of a QRM and can repeatedly generate large-scale ground-state samples.

After a sufficient number of optimization epochs, the DCGAN model successfully generates large-scale samples consistent with the true data distribution. To verify the validity of the generated data, we measure the average magnetization and potential energy density of the QRM under equivalent statistical conditions, expressed as

\begin{equation}
M=\frac{1}{N}\langle | \sum_k \hat{\mathbf{n}}_k| \rangle, {\quad}\epsilon_{\mathrm{p}}=-\frac{J}{N}\left\langle\sum_{\langle k, l\rangle} \hat{\mathbf{n}}_k \cdot \hat{\mathbf{n}}_l\right\rangle,
\end{equation}

where $M$ denotes the average magnetization of the system, $\epsilon_{\mathrm{p}}$ represents the potential energy density in the Hamiltonian, and $N$ is the total number of quantum rotors.  

Preliminary tests of accelerated large-scale equilibrium-state generation are presented in Fig.~\ref{f_6}. At $g = 4.254$, we use the two-dimensional equilibrium-state samples of the QRM with $L = 4$ as the training dataset, and the generator outputs equilibrium-state samples with $L = 8$. From 10,000 generated $(8, 8)$ matrices, we compute the variation of the average magnetization $M$ as a function of the optimization steps. As shown in Fig.~\ref{f_6}(a), although the generated samples initially deviate significantly from the true ground-state distribution, the DCGAN quickly stabilizes and achieves close agreement in the later stages of optimization. Fig.~\ref{f_6}(b) further compares the computational cost required to obtain ground-state samples of equivalent size. Under the same training conditions, the customized DCGAN achieves more than an 80\% reduction in time compared with conventional optimization-based sampling, demonstrating its efficiency and scalability for generating large-size equilibrium configurations of the QRM.

To further validate the effectiveness of the proposed DCGAN model in generating large-scale equilibrium-state samples of the QRM, we conducted additional tests on systems exhibiting different magnetic states. Specifically, we performed the same DCGAN training procedure at $g = 3.50$ and $g = 8.0$, corresponding to the ferromagnetic and paramagnetic regimes, respectively. Under equivalent statistical conditions, we measured the evolution of the average magnetization and potential energy density during the optimization process for the original $L = 8$ ground-state samples.  

As shown in Fig.~\ref{f_7}(a), the DCGAN-generated samples reproduce the average magnetization of the true ground-state configurations in both the ferromagnetic and paramagnetic phases after sufficient optimization steps. Fig.~\ref{f_7}(b) presents the evolution of the potential energy density during training, where the darker curves represent the statistical averages of the DCGAN-generated datasets. Although deviations from the ground-state values are evident during the early stages of training, the DCGAN quickly converges, achieving statistical consistency with the original samples while requiring less than one-fifth of the training time. These results confirm that the samples generated by DCGAN can accurately reproduce the equilibrium ground states of the two-dimensional QRM in different magnetic regions at the benchmark level of the statistical values of observables.

For the QRM, the computational cost associated with evaluating the many-body wavefunction increases rapidly with system size due to the growing number of coupled rotors. During numerical simulations, the sample size required to optimize the ground-state energy of the Hamiltonian also becomes statistically demanding. Nevertheless, finite-size effects play a crucial role in the study of quantum phase transitions. Large-scale sample data are essential for accurately evaluating statistical observables, identifying critical points, and determining critical exponents. Under limited computational resources, the DCGAN framework provides an effective auxiliary approach for exploring the equilibrium properties of larger quantum rotor systems, enabling the investigation of critical behavior beyond the reach of conventional numerical methods.

\section{Conclusion}

In this work, we designed two variant architectures of GANs to achieve two main objectives: (i) identifying the critical point of the two-dimensional QRM, and (ii) accelerating the generation of large-scale ground-state samples. Within a semi-supervised learning framework, we developed a Conditional GAN with a specially designed generator structure and an adaptively tuned discriminator. In the generator, both noise and test data are preprocessed and transformed through transposed convolutional layers for dimensionality reduction. After training, the encoded representations extracted from the testing dataset effectively distinguish between the quantum paramagnetic and ferromagnetic phases of the rotor model. The curvature extrema of the one-dimensional encoded data at different system sizes provide a reliable indicator of the finite-size critical points, and a linear extrapolation yields the critical coupling of the two-dimensional QRM.

To accelerate the generation of large-scale ground-state samples, we further customized a Deep Convolutional GAN architecture by introducing an upsampling process in the generator, thereby doubling the output dimension, and by defining a loss function augmented with dynamically weighted statistical descriptors of the generated distributions. The results demonstrate that after sufficient training, the ground-state samples generated by our DCGAN for the two-dimensional QRM exhibit high fidelity to benchmark observables. This approach achieves a substantial reduction in computational overhead while maintaining equivalent statistical precision. Our results indicate that the latent representations learned by generative deep learning hold broad potential for quantum phase transition problems featuring complex feature recognition, and that the generative models themselves can be purposefully designed to accelerate the production of large volumes of simulated samples.

\section{Acknowledgements}
This work is supported in part by the National Key Research and Development Program of China under Grant No. 2024YFA1611003, the Fundamental Research Funds for Central China Normal University(CCNU24JC007), and the 111 Project, with Grant No. BP0820038. Financially supported by self-determined research funds of CCNU from the college basic research and operation of MOE.


\nocite{*}

\bibliography{apssamp}

\end{document}